\documentclass[conference,letterpaper]{IEEEtran}
\IEEEoverridecommandlockouts

\usepackage[utf8]{inputenc} 
\usepackage[T1]{fontenc}
\usepackage{url}
\usepackage{ifthen}
\usepackage{cite}
\usepackage[cmex10]{amsmath} 
\usepackage{amssymb}
\usepackage{xifthen,xparse}

\usepackage{blkarray, bigstrut}
\usepackage{booktabs}

\usepackage{tikz,pgfplots,tkz-graph}
\pgfplotsset{compat=newest}
\pgfplotsset{plot coordinates/math parser=false}
\usetikzlibrary{decorations.markings,calc,positioning,matrix}

\newtheorem{theorem}{Theorem}
\newtheorem{corollary}[theorem]{Corollary}
\newtheorem{remark}[theorem]{Remark}
\newtheorem{lemma}[theorem]{Lemma}
\newtheorem{definition}[theorem]{Definition}

\newtheorem{example}{Example}

\newcommand{\ket}[1]{\left\lvert #1 \right\rangle}
\newcommand{\bra}[1]{\left\langle #1 \right\rvert}
\NewDocumentCommand\ketbra{+m+g}{%
  \IfNoValueTF{#2}
    {\left\lvert #1 \right\rangle \left\langle #1 \right\vert}
  {\left\lvert #1 \right\rangle \left\langle #2 \right\rvert}%
}
\NewDocumentCommand\braket{+m+g}{%
  \IfNoValueTF{#2}
    {\left\langle #1 \vert #1 \right\rangle}
  {\left\langle #1 \vert #2 \right\rangle}%
}

\newcommand{\vecnot}[1]{\underline{#1}}

\newcommand{\llbr}{[\![}
\newcommand{\rrbr}{]\!]}

\newcommand{\syminn}[2]{\langle #1, #2 \rangle_{\text{s}}}

\begin{document}
\title{Classical Coding Problem from Transversal $T$ Gates}

 \author{%
   \IEEEauthorblockN{Narayanan Rengaswamy\IEEEauthorrefmark{1},
                     Robert Calderbank\IEEEauthorrefmark{1},
                     Michael Newman\IEEEauthorrefmark{2},
                     and Henry D. Pfister\IEEEauthorrefmark{1}}
   \IEEEauthorblockA{\IEEEauthorrefmark{1}%
                     Department of Electrical and Computer Engineering,
                     Duke University,
                     Durham, NC 27708, USA}
    \IEEEauthorblockA{\IEEEauthorrefmark{2}%
                     Departments of Physics and Electrical and Computer Engineering,
                     Duke University,
                     Durham, NC 27708, USA\\
                     Email: \{narayanan.rengaswamy, robert.calderbank, michael.newman,  henry.pfister\}@duke.edu}
\thanks{The work of Calderbank, Rengaswamy and Pfister was supported in part by the National Science Foundation (NSF) under Grant No. 1908730. The research of Newman was supported under the ODNI/IARPA LogiQ program (W911NF-16-1-0082). Any opinions, findings, conclusions, and recommendations expressed in this material are those of the authors and do not necessarily reflect the views of these sponsors.}
%
}

\maketitle

\begin{abstract}
Universal quantum computation requires the implementation of a logical non-Clifford gate.
In this paper, we characterize all stabilizer codes whose code subspaces are preserved under physical $T$ and $T^{\dagger}$ gates.
For example, this could enable magic state distillation with non-CSS codes and, thus, provide better parameters than CSS-based protocols.
However, among non-degenerate stabilizer codes that support transversal $T$, we prove that CSS codes are optimal.
We also show that triorthogonal codes are, essentially, the only family of CSS codes that realize logical transversal $T$ via physical transversal $T$.
Using our algebraic approach, we reveal new purely-classical coding problems that are intimately related to the realization of logical operations via transversal $T$.
Decreasing monomial codes are also used to construct a code that realizes logical CCZ.
Finally, we use Ax's theorem to characterize the logical operation realized on a family of quantum Reed-Muller codes.
This result is generalized to finer angle $Z$-rotations in https://arxiv.org/abs/1910.09333.
\end{abstract}



\section{Introduction}
\label{sec:intro}

Quantum computers have been theoretically shown to provide computational advantages over conventional (classical) computers, which could have impacts in fields as varied as quantum simulation, optimization, chemistry, communications, and metrology.
Recently, Google and NASA demonstrated a computational advantage for a random circuit sampling task via a \emph{real} experiment on their 53-qubit quantum machine~\cite{Arute-nature19}.
Although the extent of the advantage has been disputed by IBM~\cite{Pednault-arxiv19}, it is widely accepted that this is a milestone hardware demonstration.
However, these computers are still very noisy and algorithms that are sensitive to noise are not within reach. 
One example is Shor's algorithm for factoring integers~\cite{Shor-focs94,Shor-siamjc97}, which has huge implications for digital security.
A quantum error correcting code (QECC) provides resilience to noise, and in this paper we focus on fault-tolerant implementation of a universal set of gates on the qubits protected by a QECC.

Universality requires one to realize a logical \emph{non-Clifford} gate and the easiest fault-tolerant realization is a \emph{transversal} operation, which splits into gates on individual qubits.
In other words, given an $\llbr n,k,d \rrbr$ QECC, we would like to understand the $k$-qubit (logical) gates that can be realized as transversal operations on the $n$ physical qubits of the code. 
Since addressing this question in full generality is challenging, in this paper we algebraically characterize all $\llbr n,k,d \rrbr$ \emph{stabilizer} QECCs~\cite{Gottesman-phd97,Calderbank-it98*2} whose code subspaces are preserved by a given pattern of $T$ and $T^{\dagger}$ gates on the $n$ qubits, i.e., this transversal operation induces some logical operation on the $k$ protected qubits.
This characterization encompasses all schemes in the literature that use transversal $T$ gates on stabilizer codes to achieve their objective. 
For example,~\cite{Bravyi-pra05,Bravyi-pra12} use this approach for \emph{magic state distillation (MSD)}.

In particular, for state distillation, almost all existing protocols use \emph{Calderbank-Shor-Steane (CSS)} codes~\cite{Calderbank-physreva96,Steane-physreva96}, which form a subclass of stabilizer codes.
Our results can be used to construct distillation protocols that utilize transversal gates on non-CSS stabilizer codes. 
At first look, this points towards the possibility of better parameters than CSS-based protocols.
However, we prove that, given any $\llbr n,k,d \rrbr$ \emph{non-degenerate} stabilizer code supporting a pattern of $T$ and $T^{\dagger}$, there exists an $\llbr n,k,d \rrbr$ CSS code with the same property.
Here, by non-degenerate we mean that each stabilizer element acts non-trivially on at least $d$ physical qubits.
While the degenerate case remains unsolved, our algebraic approach enables one to reason about CSS optimality for transversal $Z$-rotations, which is an important open problem in quantum error correction.

When our main result (Theorem~\ref{thm:transversal_T}) is specialized to CSS codes we obtain new classical coding problems, and the general case is quite similar.
Since this is a self-contained problem that classical coding theorists can analyze, we describe it here.

\textbf{CSS-T Codes:} A pair $(C_1,C_2)$ of binary linear codes with parameters $[n,k_1,d_1]$ and $[n,k_2,d_2]$, respectively, such that $C_2 \subset C_1$ and the following properties hold:
\begin{enumerate}
    \item $C_2$ is an even code, i.e., $w_H(x) \equiv 0$ (mod $2$) for all $x \in C_2$, where $w_H(x)$ is the Hamming weight of $x$.
    
    \item For each $x \in C_2$, there exists a dimension $w_H(x)/2$ self-dual code in $C_1^{\perp}$ that is supported on $x$, i.e., there exists $ C_x \subseteq C_1^{\perp}$ s.t. $|C_x| = 2^{w_H(x)/2}, C_x = C_x^{\perp}$, and $z \in C_x \Rightarrow z \preceq x$, i.e., $\text{supp}(z) \subseteq \text{supp}(x)$, where $C_1^{\perp}$ is the code dual to $C_1$ and $\text{supp}(x)$ is the support of $x$.
\end{enumerate}
\textbf{Open Problem:} A $\llbr n, k_1-k_2, \text{min}(d_1,d_2^{\perp}) \rrbr$ family of CSS-T codes such that $\frac{(k_1-k_2)}{n} = \Omega(1)$ and (ideally) $\frac{\text{min}(d_1,d_2^{\perp})}{n} = \Omega(1)$, where $d_2^{\perp}$ is the minimum distance of $C_2^{\perp}$.

This specific code family arises when the $T$ gate is applied transversally, but different patterns of $T$ and $T^{\dagger}$ gates produce variants of it~\cite{Rengaswamy-arxiv19c}.
A $\llbr 2^m, \binom{m}{m/3}, 2^{m/3} \rrbr$ Reed-Muller CSS-T family is described by $C_1 = \text{RM}(m/3,m), \ C_2 = \text{RM}(m/3-1,m)$.
However, this family has vanishing rate and distance.
It is an important open problem to construct a constant rate CSS-T family with growing distance.
For example, this would enable \emph{constant overhead} MSD, since the ratio of input noisy states to output $\epsilon$-noisy states is $O\left( \log^{\gamma}\left( \frac{1}{\epsilon} \right) \right)$, where $\gamma \triangleq \frac{\log(n/k)}{\log(d)}$ for an $\llbr n,k,d \rrbr$ code~\cite{Bravyi-pra12}.
This leads to a tremendous decrease in resource counts for this critical subroutine~\cite{Hastings-prl18}.

Several researchers have worked on constructing codes that support $T$ gates.
One of the earliest known codes to support transversal $T$ is the $\llbr 15,1,3 \rrbr$ (CSS) quantum Reed-Muller (QRM) code~\cite{Knill-arxiv96,Bravyi-pra05,Anderson-prl14}.
Subsequently, \emph{triorthogonal codes}~\cite{Bravyi-pra12} were developed to produce a systematic construction of CSS codes where the logical transversal $T$ can be realized via physical transversal $T$ (up to diagonal Clifford corrections).
In Section~\ref{sec:logical_T} we will show that this is essentially the only family of CSS codes that satisfies this property.
The topological family of \emph{3D color codes}~\cite{Kubica-pra15} has also been shown to support transversal $T$ gates.
More recently, \emph{quasitransversality}~\cite{Campbell-pra17} and the implied \emph{generalized triorthogonality}~\cite{Haah-quantum17b} conditions have been developed to construct CSS codes that support transversal $T$.
Finally, \emph{quantum pin codes}~\cite{Vuillot-arxiv19} are CSS codes that are inspired by topological codes, but they have a more general abstract construction that intrinsically supports (quasi-)transversal $Z$-rotations.

The approach in this prior work is to analyze the CSS basis states. 
Our approach is different in that we analyze the operators in the stabilizer, and it is more general, in that it extends beyond CSS codes.
For details and proofs see~\cite{Rengaswamy-arxiv19c}.


\section{Background and Notation}
\label{sec:background}

\subsection{Heisenberg-Weyl and Clifford Groups}
\label{sec:hw_clifford}

The $1$-qubit \emph{Pauli} operators are the unitaries $I_2$ (identity),
\begin{align}
X \triangleq 
\begin{bmatrix}
0 & 1 \\
1 & 0
\end{bmatrix}, \ 
Z \triangleq 
\begin{bmatrix}
1 & 0 \\
0 & -1
\end{bmatrix}, \ 
Y \triangleq \imath X Z = 
\begin{bmatrix}
0 & -\imath \\
\imath & 0
\end{bmatrix},
\end{align}
where $\imath \triangleq \sqrt{-1}$.
They satisfy $X^2 = Z^2 = Y^2 = I_2$.
The $n$-qubit \emph{Heisenberg-Weyl} (or Pauli) group $HW_N, N \triangleq 2^n$, consists of Kronecker products of these single-qubit operators with overall phases $\imath^{\kappa}, \kappa \in \mathbb{Z}_4 \triangleq \{0,1,2,3\}$.
We represent a Hermitian Pauli matrix via two binary vectors $a = [\alpha_1, \ldots, \alpha_n], b = [\beta_1, \ldots, \beta_n] \in \mathbb{Z}_2^n$ with the notation
\begin{align}
E(a,b) \triangleq \left( \imath^{\alpha_1 \beta_1} X^{\alpha_1} Z^{\beta_1} \right) \otimes \cdots \otimes \left( \imath^{\alpha_n \beta_n} X^{\alpha_n} Z^{\beta_n} \right).
\end{align}
Two Pauli matrices $E(a,b)$ and $E(c,d)$ commute if the \emph{symplectic inner product} $\syminn{[a,b]}{[c,d]} \triangleq ad^T + bc^T\ (\bmod\ 2) = 0$, and they anti-commute otherwise~\cite{Rengaswamy-isit18}.

Throughout the paper, $\oplus$ denotes modulo-$2$ addition and $+$ denotes standard integer addition.
Also, all binary and integer-valued vectors will be row vectors while complex-valued vectors will be column vectors.
For $x = [x_1,\ldots,x_n] ,y = [y_1,\ldots,y_n] \in \mathbb{Z}_2^n$, we define $x \ast y \triangleq [x_1 y_1, \ldots, x_n y_n]$. 

The \emph{Clifford} group $\text{Cliff}_N$ is the normalizer of $HW_N$ in $\mathbb{U}_N$, the unitary group of $N \times N$ matrices.
Hence, for $g \in \text{Cliff}_N$,
\begin{align}
\label{eq:cliff_action}
g E(a,b) g^{\dagger} & = \pm E([a,b] F_g), \ 
\text{where} \ F_g \Omega F_g^T = \Omega = 
{\small
\begin{bmatrix}
0 & I_n \\
I_n & 0
\end{bmatrix}}.
\end{align}
So $F_g$ is a \emph{binary symplectic matrix}, i.e., it preserves symplectic inner products $\syminn{[a,b]}{[c,d]} = [a,b]\, \Omega\, [c,d]^T$.
Since, up to scalars, $\text{Cliff}_N$ is a finite subgroup of $\mathbb{U}_N$, it is insufficient to perform \emph{universal} quantum computation.
It is well-known that $\text{Cliff}_N$ augmented by \emph{any} non-Clifford unitary can approximate any other unitary operator arbitrarily well.
A standard choice is the ``$T$'' gate $T \triangleq P^{1/2} \triangleq Z^{1/4}$~\cite{Boykin-arxiv99}.

\subsection{Quadratic Form Diagonal (QFD) Gates}
\label{sec:qfd}

The \emph{Clifford hierarchy} is a hierarchy of unitary operators first defined by Gottesman and Chuang~\cite{Gottesman-nature99} to demonstrate universal quantum computation via teleportation.
The first level of the hierarchy is $\mathcal{C}^{(1)} \triangleq HW_N$ and the subsequent levels $\ell \geq 2$ are defined recursively by
\begin{align}
\mathcal{C}^{(\ell)} \triangleq \{ U \in \mathbb{U}_N \colon U E(a,b) U^{\dagger} \in \mathcal{C}^{(\ell - 1)}\ \forall \ a,b \in \mathbb{Z}_2^n \}.
\end{align}
From this definition, it is easily seen that $\mathcal{C}^{(2)} = \text{Cliff}_N$.
Cui et al.~\cite{Cui-physreva17} described the structure of all \emph{diagonal} unitaries in this hierarchy. 
In particular, they showed that the entries in such unitaries have to be of the form $\exp\left( \frac{2\pi\imath q}{2^{\ell}} \right)$, where $q \in \mathbb{Z}_{2^{\ell}}$.

In~\cite{Rengaswamy-pra19}, the set of \emph{QFD} gates is introduced and defined by
\begin{align}
\tau_R^{(\ell)} \triangleq \sum_{v \in \mathbb{Z}_2^n} \xi^{v R v^T \bmod 2^{\ell}} \ketbra{v}, 
\end{align} 
where $\xi \triangleq \exp\left( \frac{2\pi\imath}{2^{\ell}} \right), R \in \mathbb{Z}_{2^{\ell}}^{n \times n}$ is symmetric, $\ket{v} = e_v$ is the standard basis vector in $\mathbb{C}^N$ with a $1$ in the entry indexed by $v \in \mathbb{Z}_2^n$, and $\bra{v} \triangleq \ket{v}^{\dagger}$.
It is shown that all $1$- and $2$- local diagonal gates in the hierarchy are QFD, e.g., $T = \tau_{[\, 1\, ]}^{(3)}$.
Moreover, their action on Pauli operators is characterized by
\begin{align}
\label{eq:tau_Eab}
\tau_R^{(\ell)} E(a,b) & ( \tau_R^{(\ell)} )^{\dagger} \nonumber \\
  & = \xi^{\phi(R,a,b,\ell)} E(a_0, b_0 + a_0 R) \, \tau_{\tilde{R}(R,a,\ell)}^{(\ell-1)}, \\
\phi(R,a,b,\ell) & \triangleq (1 - 2^{\ell-2}) a_0 R a_0^T + 2^{\ell-1} (a_0 b_1^T + b_0 a_1^T), \\
\tilde{R}(R,a,\ell) & \triangleq (1 + 2^{\ell-2}) D_{a_0 R} - (D_{\bar{a}_0} R D_{a_0} \nonumber \\
  & \hspace{1cm} + D_{a_0} R D_{\bar{a}_0} + 2 D_{a_0 R D_{a_0}}) \in \mathbb{Z}_{2^{\ell - 1}}^{n \times n}.
\end{align}
Equation~\eqref{eq:tau_Eab} naturally extends the action in~\eqref{eq:cliff_action} to a large class of diagonal unitaries, e.g., $TXT^{\dagger} = e^{-\imath\pi/4} Y P,\ P \triangleq \sqrt{Z}$.
Note that the symplectic matrix in this case is 
$\Gamma_R = 
\begin{bmatrix}
I_n & R \\
0 & I_n
\end{bmatrix}$ (defined over $\mathbb{Z}_{2^{\ell}}$), which also satisfies $\Gamma_R\, \Omega\, \Gamma_R^T = \Omega\ (\text{mod}\ 2)$.
Here, $D_x$ represents a diagonal matrix with the diagonal set to the vector $x$, and $\bar{x} = \vecnot{1} - x$ with $\vecnot{1}$ representing the vector whose entries are all $1$.
We write $a = a_0 + 2a_1 + 4a_2 + \ldots, b = b_0 + 2b_1 + 4b_2 + \ldots \in \mathbb{Z}^n$ with $a_i,b_i \in \mathbb{Z}_2^n$.
With this notation, $b_0 + a_0 R$ is an integer sum and the definition of $E(a,b)$ has been suitably generalized to integer vectors $a,b$ (see~\cite{Rengaswamy-pra19}).

\subsection{Stabilizer Codes}
\label{sec:stabilizer_codes}

A \emph{stabilizer group} $S$ is a commutative subgroup of $HW_N$ with Hermitian elements that does not contain $-I_N$.
If $S$ has $r$ generators, then it can be expressed as $S = \langle \nu_i E(c_i,d_i) ; i = 1,\ldots,r \rangle$, where $\nu_i \in \{ \pm 1 \}$ and $E(c_i,d_i), E(c_j,d_j)$ commute for all $i,j \in \{1,\ldots,r\}$, i.e., $\syminn{[c_i,d_i]}{[c_j,d_j]} = 0$ (mod $2$).
Given a stabilizer $S$, the associated $\llbr n,k,d \rrbr$ \emph{stabilizer code} is defined as $V(S) \triangleq \{ \ket{\psi} \in \mathbb{C}^N \colon g \ket{\psi} = \ket{\psi}\ \text{for\ all}\ g \in S \}$, where $k \triangleq n-r$ and $d$ is the \emph{distance} of the code that is defined as the minimum weight of an undetectable error.

A \emph{Calderbank-Shor-Steane (CSS)} code has a set of purely $X$-type and purely $Z$-type stabilizer generators.
Consider two classical binary codes $C_1,C_2$ such that $C_2 \subset C_1$, and let $C_1^{\perp}, C_2^{\perp}$ represent their respective dual codes.
Then, $C_1^{\perp} \subset C_2^{\perp}$ and the stabilizer for the resulting CSS code is given by $S \triangleq \langle \nu_c E(c,0), \nu_d E(0,d), c \in C_2, d \in C_1^{\perp} \rangle$ for some suitable $\nu_c, \nu_d \in \{ \pm 1 \}$.
Let $C_1$ be an $[n,k_1]$ code and $C_2$ be an $[n,k_2]$ code such that $C_1$ and $C_2^{\perp}$ can correct up to $t$ errors.
Then, $S$ defines an $\llbr n, k_1-k_2, \geq 2t+1 \rrbr$ CSS code that we will denote by CSS($X,C_2 ; Z, C_1^{\perp}$).
If $G_2$ and $G_1^{\perp}$ are generator matrices for the codes $C_2$ and $C_1^{\perp}$, respectively, then a generator matrix for the binary representation of stabilizers can be written as
\begin{align}
\setlength\aboverulesep{0pt}\setlength\belowrulesep{0pt}
    \setlength\cmidrulewidth{0.5pt}
G_S = 
\begin{blockarray}{ccc}
 n & n &   \\
\begin{block}{[c|c]c}
\hspace*{1cm} & G_1^{\perp} & n - k_1 \\
\cmidrule(lr){1-2}
G_2 & \hspace*{1cm} & k_2 \\
\end{block}
\end{blockarray}.
\end{align}

For any $S$, the projector on to the code $V(S)$ is given by
\begin{align}
\label{eq:code_projector}
\Pi_S \triangleq \prod_{i=1}^{r} \frac{\left( I_N + \nu_i E(c_i,d_i) \right)}{2} = \frac{1}{2^r} \sum_{j = 1}^{2^r} \epsilon_j E(a_j, b_j),
\end{align}
where $\epsilon_j \in \{ \pm 1 \}$ in the last equality is 
determined by the product of signs of the generators of $S$ that multiply to produce the stabilizer element $E(a_j,b_j)$.

\section{Stabilizer Codes Supporting QFD Gates}
\label{sec:stabilizer_qfd}

In order to perform universal fault-tolerant quantum computation with stabilizer QECCs, we need to identify fault-tolerant realizations of the necessary logical operators.
For logical Pauli operators, there are at least two known algorithms~\cite{Gottesman-phd97,Wilde-physreva09} to translate them into the relevant physical Pauli operators for stabilizer codes.
At the second level of the Clifford hierarchy, for logical Clifford gates, there have been several works that determine fault-tolerant realizations on specific codes or code families.
In~\cite{Rengaswamy-isit18,Rengaswamy-arxiv19b} we developed a systematic and efficient algorithm using symplectic matrices to translate logical Clifford circuits into physical Clifford circuits for \emph{any} stabilizer code.
Although this \emph{Logical Clifford Synthesis (LCS)} algorithm currently does not guarantee fault-tolerance of the solutions, a better understanding of the symplectic solution space might help us achieve that objective. 

For non-Clifford gates, the lack of a symplectic formalism and the fact that Paulis are not mapped to Paulis under conjugation together make synthesis of logical non-Clifford gates much harder.
Therefore, our first goal is to understand the structure required in the stabilizer so that a specified (non-Clifford) gate preserves the code subspace.
In this paper we restrict ourselves to physical QFD gates since we have an extension of the symplectic formalism for these gates.
We will discuss two steps involved in achieving this goal and solve the transversal $T$ special case completely.
For proofs, refer to~\cite{Rengaswamy-arxiv19c}. 

\vspace{0.075cm}

\noindent\textbf{Step 1:} Express QFD action on Pauli matrices in Pauli basis.

\vspace{0.075cm}

First we expand $\tau_R^{(\ell)} = \sum_{x \in \mathbb{Z}_2^n} c_{R,x}^{(\ell)} \cdot \frac{1}{\sqrt{2^n}} E(0,x)$, where
\begin{align}
\label{eq:tau_coeff}
c_{R,x}^{(\ell)} \triangleq \text{Tr}\left[ \frac{E(0,x)}{\sqrt{2^n}} \tau_R^{(\ell)} \right] = \frac{1}{\sqrt{2^n}} \sum_{v \in \mathbb{Z}_2^n} (-1)^{vx^T} \xi^{vRv^T}. 
\end{align}
Applying this for $\tau_{\tilde{R}(R,a,\ell)}^{(\ell-1)}$ in~\eqref{eq:tau_Eab} we get, assuming $a,b \in \mathbb{Z}_2^n$,
\begin{align}
& \tau_R^{(\ell)} E(a,b) ( \tau_R^{(\ell)} )^{\dagger} \nonumber \\
  & = \xi^{\phi(R,a,b,\ell)} E(a, b + a R) \, \tau_{\tilde{R}(R,a,\ell)}^{(\ell-1)} \nonumber \\
  & = \xi^{\phi(R,a,b,\ell)} E(a, b + a R) \cdot \frac{1}{\sqrt{2^n}} \sum_{x \in \mathbb{Z}_2^n} c_{\tilde{R}(R,a,\ell),x}^{(\ell-1)} E(0,x) \nonumber \\
\label{eq:tau_Eab_expand}
  & = \frac{\xi^{\phi(R,a,b,\ell)}}{\sqrt{2^n}} \sum_{x \in \mathbb{Z}_2^n} c_{\tilde{R}(R,a,\ell),x}^{(\ell-1)} \imath^{-ax^T} E(a, b + aR + x).
\end{align}
The primary problem here is to determine which coefficients are non-zero for given $R,a,\ell$, and to compute their values.

\begin{lemma}
\label{lem:conj_by_trans_T}
Let $E(a,b) \in HW_N$, for some $a,b \in \mathbb{Z}_2^n$.
Then the transversal $T$ gate acts on $E(a,b)$ as
\begin{align*}
T^{\otimes n} E(a,b) \left( T^{\otimes n} \right)^{\dagger} = \frac{1}{2^{w_H(a)/2}} \sum_{y \preceq a} (-1)^{b y^T} E(a, b \oplus y),
\end{align*}
where $w_H(a) = aa^T$ is the Hamming weight of $a$, and $y \preceq a$ denotes that support of $y$ is contained in the support of $a$.
\end{lemma}

For the general case where each qubit is acted upon by a possibly different integer power of $T$, we provide the result in~\cite{Rengaswamy-arxiv19c}.
These formulae may be of independent interest. 

\vspace{0.075cm}

\noindent \textbf{Step 2:} Determine conditions on $S$ for $\tau_R^{(\ell)} \Pi_S (\tau_R^{(\ell)})^{\dagger} = \Pi_S$.

\vspace{0.075cm}

We focus on the above equality because this is the necessary and sufficient condition for a (QFD) unitary to preserve the code subspace (see~\cite{Rengaswamy-arxiv19c} for a simple argument).
By expanding the above equality for $T^{\otimes n}$ using the result in Step 1, we get
\begin{align}
& T^{\otimes n} \Pi_S \left( T^{\otimes n} \right)^{\dagger} \nonumber \\
  & = \frac{1}{2^r} \sum_{j = 1}^{2^r} \epsilon_j \left[ T^{\otimes n} E(a_j, b_j) \left( T^{\otimes n} \right)^{\dagger} \right] \\
  & = \frac{1}{2^r} \sum_{j = 1}^{2^r} \frac{\epsilon_j}{2^{w_H(a_j)/2}} \sum_{y \preceq a_j} (-1)^{b_j y^T} E(a_j, b_j \oplus y).
%
\end{align}
This needs to equal~\eqref{eq:code_projector} and the following characterizes that.

\begin{theorem}
\label{thm:transversal_T}
%
%
%
%
Let $S = \langle \nu_i E(c_i,d_i) ; i = 1,\ldots,r \rangle$ define a stabilizer code, with arbitrary $\nu_i \in \{ \pm 1 \}$, and denote the elements of $S$ by $\epsilon_j E(a_j, b_j), j = 1,2,\ldots,2^r$. 
If the transversal application of the $T$ gate  preserves the code space $V(S)$ and hence realizes a logical operation on $V(S)$, then:
\begin{enumerate}

\item For any $\epsilon_j E(a_j,b_j) \in S$, $w_H(a_j)$ is even, where $w_H(a_j)$ represents the Hamming weight of $a_j \in \mathbb{Z}_2^n$.

\item  For any $\epsilon_j E(a_j,b_j) \in S$ with non-zero $a_j$, define $Z_j \triangleq \{ z \preceq a_j \colon \epsilon_z E(0,z) \in S\ \text{for\ some}\ \epsilon_z \in \{ \pm 1 \} \}$.%
      Then $Z_j$ contains its dual computed only on the support of $a_j$, i.e., on the ambient dimension $w_H(a_j)$.
      Equivalently, $Z_j$ contains a dimension $w_H(a_j)/2$ self-dual code $A_j$ that is supported on $a_j$, i.e., there exists a subspace $A_j \subseteq Z_j$ such that $y z^T = 0$ (mod 2) for any $y,z \in A_j$ (including $y = z$) and $\text{dim}(A_j) = w_H(a_j)/2$.
      
\item Let $\tilde{Z}_j \subseteq \mathbb{Z}_2^{w_H(a_j)}$ represent $Z_j$ with all positions outside the support of $a_j$ punctured (dropped).
      Then, for each $z \in \mathbb{Z}_2^n$ such that $\tilde{z} \in (\tilde{Z}_j)^{\perp}$ for some $j \in \{ 1,\ldots,2^r \}$, we have $\epsilon_z = \imath^{zz^T}$, i.e., $\imath^{zz^T} E(0,z) \in S$.
      Here, $(\tilde{Z}_j)^{\perp}$ denotes the dual of $Z_j$ taken over this punctured space with ambient dimension $w_H(a_j)$.
      (Also, $Z_j \supseteq (\tilde{Z}_j)^{\perp}$ with zeros added outside the support of $a_j$.)

\end{enumerate} 
Conversely, if the first two conditions above are satisfied, and if the third condition holds for all $z \in A_j$ instead of just the dual of (the punctured) $Z_j$, then transversal $T$ preserves the code space $V(S)$ and hence induces a logical operation.
\end{theorem}
We will illustrate this theorem using a simple CSS example.

\begin{example}
\label{eg:cssT_622}
\normalfont
Define a $\llbr 6,2,2 \rrbr$ CSS code by the matrix
\begin{align}
\setlength\aboverulesep{0pt}\setlength\belowrulesep{0pt}
    \setlength\cmidrulewidth{0.5pt}
G_S = 
\left[
\begin{array}{cccccc|cccccc}
1 & 1 & 1 & 1 & 1 & 1 & 0 & 0 & 0 & 0 & 0 & 0 \\
\hline 
0 & 0 & 0 & 0 & 0 & 0 & 1 & 1 & 0 & 0 & 0 & 0 \\
0 & 0 & 0 & 0 & 0 & 0 & 0 & 0 & 1 & 1 & 0 & 0 \\
0 & 0 & 0 & 0 & 0 & 0 & 0 & 0 & 0 & 0 & 1 & 1 \\
\end{array}
\right].
\end{align}
The right half of the last $3$ rows form the generators of $Z_S$ for this code.
Since there is only one non-trivial $a_j$ in this case, we see that $Z_S = A_1$ with $a_1 = [1,1,1,1,1,1]$.
Hence, the stabilizer generators are $X^{\otimes 6} = X_1 X_2 \cdots X_6, - Z_1 Z_2, - Z_3 Z_4, - Z_5 Z_6$, since the generators of $Z_S$ have weight $2$.
Multiplying $X^{\otimes 6}$ and the product of these three $Z$-stabilizers, we see that $Y^{\otimes 6} \in S$.

We can define the logical $X$ operators for this code to be $\bar{X}_1 = X_1 X_2, \bar{X}_2 = X_3 X_4$, since these are linearly independent and commute with all stabilizers.
Then we observe 
\begin{align}
T^{\otimes 6} X_1 X_2 (T^{\otimes 6})^{\dagger} & = e^{-\imath \cdot 2\pi/4} (Y_1 P_1) (Y_2 P_2) \\
  & = -\imath \cdot (\imath X_1 Z_1 P_1) (\imath X_2 Z_2 P_2) \\
  & \equiv -\imath (X_1 X_2) (P_1 P_2),
\end{align}
since $-Z_1 Z_2 \in S$.
We observe that $(P_1 P_2) X^{\otimes 6} (P_1 P_2)^{\dagger} = Y_1 Y_2 X_3 X_4 X_5 X_6 \equiv X^{\otimes 6}$ up to the stabilizer $-Z_1 Z_2$, so $P_1 P_2$ indeed preserves $V(S)$.
But $(P_1 P_2) (X_1 X_2) (P_1 P_2)^{\dagger} = Y_1 Y_2 = (X_1 X_2) (-Z_1 Z_2) \equiv X_1 X_2$, and $P_1 P_2$ obviously commutes with $\bar{X}_2$, so $P_1 P_2$ is essentially the logical identity gate.
A similar reasoning holds for $P_3 P_4$.
Therefore, up to a global phase, the transversal $T$ preserves the logical operators $\bar{X}_1$ and $\bar{X}_2$, so in this case the transversal $T$ gate realizes just the logical identity (up to a global phase).
This can also be checked explicitly by writing the logical basis states. 

Given that $S$ has the necessary structure given by Theorem~\ref{thm:transversal_T}, note that we can freely add another $Z$-stabilizer generator that commutes with $X^{\otimes 6}$, e.g., $Z_1 Z_3 Z_4 Z_6 \leftrightarrow [1,0,1,1,0,1] \notin Z_S$.
This preserves the transversal $T$ property: once $T^{\otimes n} \Pi_S (T^{\otimes n})^{\dagger} = \Pi_S$, mapping $\Pi_S \mapsto \Pi_S \cdot \frac{(I_N + E(0,z))}{2}$ preserves equality since $E(0,z)$ is diagonal. \hfill \IEEEQEDhere
\end{example}

This example also illustrates the calculations required to determine the logical operator induced by transversal $T$ on the code space. 
Our general results in~\cite{Rengaswamy-arxiv19c} follows this strategy.


When we specialize this theorem to CSS codes, we obtain the \emph{CSS-T codes} introduced in Section~\ref{sec:intro}. 
We observe that the case of general stabilizer codes is quite similar. 
The generalization to arbitrary patterns of $T$ and $T^{\dagger}$ is given in~\cite{Rengaswamy-arxiv19c}, together with a partial extension to finer angle $Z$-rotations, which involves trigonometric quantities.

\begin{remark}
\normalfont
Intuitively, a CSS-T code is determined by two classical codes $C_2 \subset C_1$ such that for every codeword $x \in C_2$, there exists a dimension $w_H(x)/2$ self-dual code in $C_1^{\perp}$ supported on $x$.
This also means that $C_1 \ast C_2 \subseteq C_1^{\perp}$ for the following reason.
Let $a \in C_1, x \in C_2$, so that $a$ is orthogonal to every vector in $C_1^{\perp}$.
In particular, $a$ is orthogonal to the self-dual code $C_x \subset C_1^{\perp}$ supported on $x$.
But for any $z \in C_x$, $az^T = (a \ast x)z^T = 0$.
This means $a \ast x \in C_x \subset C_1^{\perp}$ since $C_x$ is self-dual.
We believe this observation can make it convenient to derive properties of CSS-T codes, e.g., using~\cite{Randriambololona-aagct15}. \hfill \IEEEQEDhere
\end{remark}

Now we provide an important corollary (see~\cite{Rengaswamy-arxiv19c} for proof).

\begin{definition}
An $\llbr n,k,d \rrbr$ stabilizer code is \emph{non-degenerate} if every stabilizer element has weight at least $d$.
\end{definition}

\begin{corollary}
\label{cor:csst_sufficient}
Consider an $\llbr n, k, d \rrbr$ non-degenerate stabilizer code $V(S)$ that satisfies Theorem~\ref{thm:transversal_T}. 
The stabilizer $S$ has generators of the form $\epsilon E(a,b), \epsilon' E(a',0), \epsilon'' E(0,b')$.
Then the $\llbr n, k, d \rrbr$ CSS code defined by replacing $\epsilon E(a,b)$'s ($a,b \neq 0$) with $\epsilon E(a,0)$'s also satisfies the transversal $T$ property, i.e., generators $\epsilon' E(a',0), \epsilon'' E(0,b')$ of $S$ are left unchanged. \hfill \IEEEQEDhere
\end{corollary}

This corollary shows that, for the purpose of transversal $T$ on non-degenerate stabilizer codes, CSS-T codes are optimal (in terms of $n,k,d$).
Therefore, magic state distillation protocols based on these codes might be nearly optimal, unless the degenerate case fails non-trivially.
We provide a brief discussion of the degenerate case in~\cite{Rengaswamy-arxiv19c}, where we show that we can extend the above corollary under an additional condition on the stabilizer of the degenerate code.

\subsection{Logical $T$ Gates from Transversal $T$}
\label{sec:logical_T}

In~\cite{Rengaswamy-arxiv19c} we revisit the well-known $\llbr 15,1,3 \rrbr$ code using classical codes and show that it is indeed a CSS-T code.
More generally, we can construct CSS-T codes where the physical transversal $T$ realizes logical transversal $T$.
In fact, \emph{triorthogonal codes} introduced by Bravyi and Haah~\cite{Bravyi-pra12} serve exactly this purpose.
As our next result, using our methods we show a ``converse'' that triorthogonality is not only sufficient but also necessary if we desire to realize logical transversal $T$ via physical transversal $T$ (using a CSS-T code).

\begin{definition}[Triorthogonality~\cite{Bravyi-pra12}]
\label{def:triorthogonality}
A $p \times q$ binary matrix $G$ is said to be \emph{triorthogonal} if and only if the support of any pair and triple of its rows has an even weight overlap, i.e., $w_H(G_a \ast G_b) \equiv 0$ (mod $2$) for any two rows $G_a$ and $G_b$ for $1 \leq a < b \leq p$, and $w_H(G_a \ast G_b \ast G_c) \equiv 0$ (mod $2$) for all triples of rows $G_a, G_b, G_c$ for $1 \leq a < b < c \leq p$.
\end{definition}

\begin{theorem}
\label{thm:logical_trans_T}
Let $S$ be the stabilizer for an $\llbr n,k,d \rrbr$ CSS-T code CSS($X, C_2 ; Z, C_1^{\perp}$).
Let $G_1 = 
\begin{bmatrix}
G_{C_1/C_2} \\
G_2
\end{bmatrix}$ be a generator matrix for the classical code $C_1 \supset C_2$ such that the rows $x_i, i = 1,\ldots,k$, of $G_{C_1/C_2}$ form a generating set for the coset space $C_1/C_2$ that produces the logical $X$ group of the CSS-T code, i.e., $\bar{X} = \langle E(x_i,0) ; i = 1,\ldots,k \rangle$.
Then physical transversal $T$ realizes logical transversal $T$, without Clifford corrections as in~\cite{Bravyi-pra12}, if and only if the matrix $G_1$ is triorthogonal and the following condition holds for all $a \in C_2$:
\begin{align*}
x = \bigoplus_{i=1}^{k} c_i x_i,\ c_i \in \{0,1\} \Rightarrow w_H(x \oplus a) \equiv w_H(c) \ (\bmod\ 8).
\end{align*}
\end{theorem}

\begin{corollary}
\label{cor:triortho_general}
The triorthogonal construction introduced by Bravyi and Haah~\cite{Bravyi-pra12} is the most general CSS family that realizes logical transversal $T$ from physical transversal $T$.
\begin{IEEEproof}
The strategy is to show that the weight condition in Theorem~\ref{thm:logical_trans_T} is equivalent to the condition one obtains by setting the Clifford correction in~\cite{Bravyi-pra12} to be trivial (see~\cite{Rengaswamy-arxiv19c}). 
\end{IEEEproof}
\end{corollary}

Note that if the weight condition in Theorem~\ref{thm:logical_trans_T} is replaced by the condition that $E(x,0) \in \bar{X} \Rightarrow \imath^{w_H(x)} E(0,x) \in S$, then the induced logical operator is trivial, i.e., the logical identity~\cite{Rengaswamy-arxiv19c}.
Since for CSS-T codes we already have $C_2 \subseteq C_1^{\perp}$, this condition is equivalent to the constraint $C_1 \subseteq C_1^{\perp}$.

\subsection{Logical Controlled-Controlled-Z Gates from Transversal $T$}
\label{sec:logical_ccz}

The gate $\text{CCZ} \triangleq \text{diag}(1,1,1,1,1,1,1,-1)$ belongs to $\mathcal{C}^{(3)}$ and enables universal computation when combined with $\mathcal{C}^{(2)}$.
One of the simplest codes that realizes logical CCZ from physical transversal $T$ is Campbell's $\llbr 8,3,2 \rrbr$ (CSS) ``smallest interesting color code''~\cite{Campbell-blog16}.
In our notation, this code is described by setting $C_2$ to be the $8$-bit repetition code $\text{RM}(0,3)$ and $C_1 = C_1^{\perp}$ to be the $[8,4,4]$ extended Hamming code, which is also the self-dual Reed-Muller code $\text{RM}(1,3)$.

A general class of polynomial evaluation codes, called \emph{decreasing monomial codes (DMCs)}, were introduced by Bardet et al.~\cite{Bardet-isit16}.
While a Reed-Muller code $\text{RM}(r,m)$ is generated by all binary $m$-variate monomials of degree up to $r \leq m$, DMCs allow one to include all monomials up to degree $r-1$ and a subset of degree-$r$ monomials according to a partial order.
This provides greater design freedom, and we refer to~\cite{Bardet-isit16} for a description of some code properties.

\begin{example}
Recently, Krishna and Tillich used DMCs to construct triorthogonal codes from punctured polar codes for magic state distillation~\cite{Krishna-arxiv18}.
We are able to construct a $\llbr 16,3,2 \rrbr$ CSS code from DMCs where transversal $T$ realizes logical CCZ.
Define the code $C_2$ as the space generated by the monomials $G_2 = \{1, x_1, x_2\}$, and the code $C_1$ as the space generated by $G_1 = G_2 \cup \{ x_3, x_4, x_1 x_2 \}$.
Hence, the logical $X$ group is generated by $G_X = \{ x_3, x_4, x_1 x_2 \}$.
Using~\cite{Bardet-isit16} it is easy to see that $G_1^{\perp} = \{ 1,x_1,x_2,x_3,x_4,x_1 x_2,x_1 x_3,x_1 x_4,x_2 x_3,x_2 x_4 \}$ and $G_2^{\perp} = G_1^{\perp} \cup \{ x_3 x_4,x_1 x_2 x_3,x_1 x_2 x_4 \}$. 
So the logical $Z$ group is generated by $G_Z = \{ x_1 x_2 x_4, x_1 x_2 x_3, x_3 x_4 \}$.

To see that this code satisfies Theorem~\ref{thm:transversal_T}, consider for example the $X$-stabilizer corresponding to the monomial $x_1 \in G_2$. 
We observe that the elements $x_1, x_1 x_2, x_1 x_3, x_1 x_4 \in G_1^{\perp}$ are supported on $x_1$.
When we project down to $x_1$, we get the monomials $1, \tilde{x}_1 = x_2, \tilde{x}_2 = x_3, \tilde{x}_3 = x_4$ that precisely generate the code RM($1,3$) that is self-dual.
A similar analysis can be made for other elements in $C_2$.
Moreover, since the elements in $G_1^{\perp}$ have weights $4,8$, or $16$, the last condition of Theorem~\ref{thm:transversal_T} does not introduce any negative signs for the $Z$-stabilizers.
We believe this is not just one special case but points towards using this formalism for a general construction of CSS codes that support transversal $Z$-rotations. 
In~\cite{Rengaswamy-arxiv19c} we also discuss connections to \emph{pin codes}~\cite{Vuillot-arxiv19}, \emph{quasitransversality}~\cite{Campbell-pra17} and the \emph{generalized triorthogonality}~\cite{Haah-quantum17b} conditions for CSS codes to realize logical CCZs from transversal $T$. \hfill \IEEEQEDhere
\end{example}

Finally we describe a $\llbr 2^m, \binom{m}{r}, 2^r \rrbr$ quantum Reed-Muller (QRM) family that we generalize to support transversal \emph{finer angle} $Z$-rotations in~\cite{Rengaswamy-arxiv19c}. 
We also characterize the exact induced logical operation through Ax's theorem on residue weights of polynomials~\cite{Ax-ajm64}.
For the $T$ case, QRM($r,m$) is described by $C_1 = \text{RM}(r,m)$ and $C_2 = \text{RM}(r-1,m)$, where $\frac{m-1}{3} < r \leq \frac{m}{3}$ ensures that transversal $T$ preserves the code space and induces a non-trivial logical gate.
This has close connections to~\cite{Haah-quantum17b}.
The argument to show that QRM($r,m$) satisfies Theorem~\ref{thm:transversal_T} is very similar to the $\llbr 16,3,2 \rrbr$ example.

\begin{example}
We use the $\llbr 64,15,4 \rrbr$ code to demonstrate the general form of the logical operation.
Here, the logical qubits $v_f \in \mathbb{Z}_2^{15}$ are identified with the degree $r=2$ monomials that define generators for logical $X$ operators.
Hence, we have
\begin{align}
\ket{v_f}_L = \ket{v_{x_1 x_2}}_L \otimes \ket{v_{x_1 x_3}}_L \otimes \cdots \otimes \ket{v_{x_5 x_6}}_L \in \mathbb{C}^{2^{15}}.
\end{align}
(The $f$ will be clarified shortly.)
The logical gate induced by $T^{\otimes 64}$ is described by $U^L \ket{v_f}_L = (-1)^{q(v_f)} \ket{v_f}_L, q(v_f) =$
\begin{align*}
     & v_{x_1 x_2} v_{x_3 x_4} v_{x_5 x_6} + v_{x_1 x_2} v_{x_3 x_5} v_{x_4 x_6} + v_{x_1 x_2} v_{x_3 x_6} v_{x_4 x_5} \nonumber \\
     + \ & v_{x_1 x_3} v_{x_2 x_4} v_{x_5 x_6} + v_{x_1 x_3} v_{x_2 x_5} v_{x_4 x_6} + v_{x_1 x_3} v_{x_2 x_6} v_{x_4 x_5}  \nonumber \\
     + \ & v_{x_1 x_4} v_{x_2 x_3} v_{x_5 x_6} + v_{x_1 x_4} v_{x_2 x_5} v_{x_3 x_6} + v_{x_1 x_4} v_{x_2 x_6} v_{x_3 x_5}  \nonumber \\
     + \ & v_{x_1 x_5} v_{x_2 x_3} v_{x_4 x_6} + v_{x_1 x_5} v_{x_2 x_4} v_{x_3 x_6} + v_{x_1 x_5} v_{x_2 x_6} v_{x_3 x_4}  \nonumber \\
     + \ & v_{x_1 x_6} v_{x_2 x_3} v_{x_4 x_5} + v_{x_1 x_6} v_{x_2 x_4} v_{x_3 x_5} + v_{x_1 x_6} v_{x_2 x_5} v_{x_3 x_4},
\end{align*}
where each term in the polynomial corresponds to a logical CCZ gate acting on the three logical qubits indexed by the three monomial subscripts, and the sum corresponds to a product of such gates (in the logical unitary space).

Recall that for $v_f \in \mathbb{Z}_2^{15}$ the CSS basis states are given by
\begin{align}
\label{eq:css_basis_states}
\ket{v_f}_L & \equiv \frac{1}{|C_2|} \sum_{c \in C_2}  \ket{v_f \cdot G_{C_1/C_2} \oplus c}. 
%
\end{align}
For QRM$(r,m)$, the rows of $G_{C_1/C_2}$ correspond to degree $r$ monomials, each identifying a logical qubit.
So a non-trivial logical $X$ operator is described by a degree $r$ polynomial $f$, but only the degree $r$ terms determine which logical qubits are acted upon.
This implies that each degree $r$ term in $f$ sets the corresponding logical qubit to $\ket{1}_L$ ($\ket{v_f}_L = \ket{0}_L$ initially).

For this code, the rows of $G_{C_1/C_2}$ are evaluations of the $15$ degree $2$ monomials, namely $x_1 x_2, x_1 x_3, x_1 x_4, \ldots, x_5 x_6$.
So the polynomial $f \in \text{RM}(r,m)$ above is a linear combination of degree $r=2$ monomials, and possibly lower degree monomials (that correspond to just $X$-type stabilizers). 
Hence, $v_f \in \mathbb{Z}_2^{15}$ exactly describes which corresponding rows of $G_{C_1/C_2}$ are chosen in this linear combination.
Therefore, if $f = x_1 x_2 + x_3 x_4 + x_5 x_6 + (\text{smaller\ degree\ terms})$, then $v_{x_1 x_2} = v_{x_3 x_4} = v_{x_5 x_6} = 1$ and other logical qubits are set to $\ket{0}_L$, so $q(v_f) = 1$.
But if $f = x_1 x_2 + x_3 x_4 + x_5 x_6 + x_3 x_5 + x_4 x_6 + (\text{smaller\ degree\ terms})$, then $q(v_f) = 0$ as this $f$ corresponds to two CCZs applying the phase $-1$.  \hfill \IEEEQEDhere
\end{example}

\IEEEtriggeratref{14}


\begin{thebibliography}{10}
\providecommand{\url}[1]{#1}
\csname url@samestyle\endcsname
\providecommand{\newblock}{\relax}
\providecommand{\bibinfo}[2]{#2}
\providecommand{\BIBentrySTDinterwordspacing}{\spaceskip=0pt\relax}
\providecommand{\BIBentryALTinterwordstretchfactor}{4}
\providecommand{\BIBentryALTinterwordspacing}{\spaceskip=\fontdimen2\font plus
\BIBentryALTinterwordstretchfactor\fontdimen3\font minus
  \fontdimen4\font\relax}
\providecommand{\BIBforeignlanguage}[2]{{%
\expandafter\ifx\csname l@#1\endcsname\relax
\typeout{** WARNING: IEEEtran.bst: No hyphenation pattern has been}%
\typeout{** loaded for the language `#1'. Using the pattern for}%
\typeout{** the default language instead.}%
\else
\language=\csname l@#1\endcsname
\fi
#2}}
\providecommand{\BIBdecl}{\relax}
\BIBdecl

\bibitem{Arute-nature19}
\BIBentryALTinterwordspacing
F.~Arute, K.~Arya, R.~Babbush, D.~Bacon, J.~C. Bardin, R.~Barends, R.~Biswas,
  S.~Boixo, F.~G. S.~L. Brandao, D.~A. Buell, B.~Burkett, Y.~Chen, Z.~Chen,
  B.~Chiaro, R.~Collins, W.~Courtney, A.~Dunsworth, E.~Farhi, B.~Foxen,
  A.~Fowler, C.~Gidney, M.~Giustina, R.~Graff, K.~Guerin, S.~Habegger, M.~P.
  Harrigan, M.~J. Hartmann, A.~Ho, M.~Hoffmann, T.~Huang, T.~S. Humble, S.~V.
  Isakov, E.~Jeffrey, Z.~Jiang, D.~Kafri, K.~Kechedzhi, J.~Kelly, P.~V. Klimov,
  S.~Knysh, A.~Korotkov, F.~Kostritsa, D.~Landhuis, M.~Lindmark, E.~Lucero,
  D.~Lyakh, S.~Mandr{\`{a}}, J.~R. McClean, M.~McEwen, A.~Megrant, X.~Mi,
  K.~Michielsen, M.~Mohseni, J.~Mutus, O.~Naaman, M.~Neeley, C.~Neill, M.~Y.
  Niu, E.~Ostby, A.~Petukhov, J.~C. Platt, C.~Quintana, E.~G. Rieffel,
  P.~Roushan, N.~C. Rubin, D.~Sank, K.~J. Satzinger, V.~Smelyanskiy, K.~J.
  Sung, M.~D. Trevithick, A.~Vainsencher, B.~Villalonga, T.~White, Z.~J. Yao,
  P.~Yeh, A.~Zalcman, H.~Neven, and J.~M. Martinis, ``{Quantum supremacy using
  a programmable superconducting processor},'' \emph{Nature}, vol. 574, no.
  7779, pp. 505--510, 2019. [Online]. Available:
  \url{http://www.nature.com/articles/s41586-019-1666-5}
\BIBentrySTDinterwordspacing

\bibitem{Pednault-arxiv19}
\BIBentryALTinterwordspacing
E.~Pednault, J.~A. Gunnels, G.~Nannicini, L.~Horesh, and R.~Wisnieff,
  ``{Leveraging Secondary Storage to Simulate Deep 54-qubit Sycamore
  Circuits},'' \emph{arXiv preprint arXiv:1910.09534}, 2019. [Online].
  Available: \url{http://arxiv.org/abs/1910.09534}
\BIBentrySTDinterwordspacing

\bibitem{Shor-focs94}
\BIBentryALTinterwordspacing
P.~Shor, ``{Algorithms for quantum computation: discrete logarithms and
  factoring},'' in \emph{Proc.\ IEEE Symp.\ on the Found.\ of Comp.\
  Sci.}\hskip 1em plus 0.5em minus 0.4em\relax IEEE Comput. Soc. Press, 1994,
  pp. 124--134. [Online]. Available:
  \url{http://ieeexplore.ieee.org/document/365700/}
\BIBentrySTDinterwordspacing

\bibitem{Shor-siamjc97}
P.~W. Shor, ``{Polynomial-Time Algorithms for Prime Factorization and Discrete
  Logarithms on a Quantum Computer},'' \emph{SIAM J.\ Comp.}, vol.~26, no.~5,
  pp. 1484--1509, 1997, [Online]. Available:
  http://arxiv.org/abs/quant-ph/9508027.

\bibitem{Gottesman-phd97}
\BIBentryALTinterwordspacing
D.~Gottesman, ``Stabilizer codes and quantum error correction,'' Ph.D.
  dissertation, California Institute of Technology, 1997. [Online]. Available:
  \url{https://arxiv.org/abs/quant-ph/9705052}
\BIBentrySTDinterwordspacing

\bibitem{Calderbank-it98*2}
\BIBentryALTinterwordspacing
R.~Calderbank, E.~Rains, P.~Shor, and N.~Sloane, ``Quantum error correction via
  codes over {GF}(4),'' \emph{IEEE Trans.\ Inform.\ Theory}, vol.~44, no.~4,
  pp. 1369--1387, Jul 1998. [Online]. Available:
  \url{https://arxiv.org/abs/quant-ph/9608006}
\BIBentrySTDinterwordspacing

\bibitem{Bravyi-pra05}
\BIBentryALTinterwordspacing
S.~Bravyi and A.~Kitaev, ``{Universal quantum computation with ideal Clifford
  gates and noisy ancillas},'' \emph{Phys. Rev. A}, vol.~71, no.~2, p. 022316,
  2005. [Online]. Available: \url{https://arxiv.org/abs/quant-ph/0403025}
\BIBentrySTDinterwordspacing

\bibitem{Bravyi-pra12}
\BIBentryALTinterwordspacing
S.~Bravyi and J.~Haah, ``{Magic-state distillation with low overhead},''
  \emph{Phys. Rev. A}, vol.~86, no.~5, p. 052329, 2012. [Online]. Available:
  \url{http://arxiv.org/abs/1209.2426}
\BIBentrySTDinterwordspacing

\bibitem{Calderbank-physreva96}
A.~R. Calderbank and P.~W. Shor, ``Good quantum error-correcting codes exist,''
  \emph{Phys. Rev. A}, vol.~54, pp. 1098--1105, Aug 1996.

\bibitem{Steane-physreva96}
A.~M. Steane, ``{Simple quantum error-correcting codes},'' \emph{Phys. Rev. A},
  vol.~54, no.~6, pp. 4741--4751, 1996.

\bibitem{Rengaswamy-arxiv19c}
\BIBentryALTinterwordspacing
N.~Rengaswamy, R.~Calderbank, M.~Newman, and H.~D. Pfister, ``{On Optimality of
  CSS Codes for Transversal $T$},'' \emph{arXiv preprint arXiv:1910.09333},
  2019. [Online]. Available: \url{http://arxiv.org/abs/1910.09333}
\BIBentrySTDinterwordspacing

\bibitem{Hastings-prl18}
\BIBentryALTinterwordspacing
M.~B. Hastings and J.~Haah, ``{Distillation with Sublogarithmic Overhead},''
  \emph{Phys. Rev. Lett.}, vol. 120, no.~5, p. 050504, 2018. [Online].
  Available: \url{https://link.aps.org/doi/10.1103/PhysRevLett.120.050504}
\BIBentrySTDinterwordspacing

\bibitem{Knill-arxiv96}
\BIBentryALTinterwordspacing
E.~Knill, R.~Laflamme, and W.~Zurek, ``{Threshold Accuracy for Quantum
  Computation},'' \emph{arXiv preprint arXiv:quant-ph/9610011}, 1996. [Online].
  Available: \url{http://arxiv.org/abs/quant-ph/9610011}
\BIBentrySTDinterwordspacing

\bibitem{Anderson-prl14}
J.~T. Anderson, G.~Duclos-Cianci, and D.~Poulin, ``{Fault-Tolerant Conversion
  between the Steane and Reed-Muller Quantum Codes},'' \emph{Phys. Rev. Lett.},
  vol. 113, no.~8, p. 080501, 2014, [Online]. Available:
  http://arxiv.org/abs/1403.2734.

\bibitem{Kubica-pra15}
\BIBentryALTinterwordspacing
A.~Kubica and M.~E. Beverland, ``{Universal transversal gates with color codes:
  A simplified approach},'' \emph{Phys. Rev. A}, vol.~91, no.~3, p. 032330,
  2015. [Online]. Available: \url{https://arxiv.org/abs/1410.0069}
\BIBentrySTDinterwordspacing

\bibitem{Campbell-pra17}
\BIBentryALTinterwordspacing
E.~T. Campbell and M.~Howard, ``{Unified framework for magic state distillation
  and multiqubit gate synthesis with reduced resource cost},'' \emph{Phys. Rev.
  A}, vol.~95, no.~2, p. 022316, Feb 2017. [Online]. Available:
  \url{https://journals.aps.org/pra/pdf/10.1103/PhysRevA.95.022316}
\BIBentrySTDinterwordspacing

\bibitem{Haah-quantum17b}
\BIBentryALTinterwordspacing
J.~Haah and M.~B. Hastings, ``{Codes and Protocols for Distilling {\$}T{\$},
  controlled-{\$}S{\$}, and Toffoli Gates},'' \emph{Quantum}, vol.~2, p.~71,
  2017. [Online]. Available: \url{https://arxiv.org/abs/1709.02832}
\BIBentrySTDinterwordspacing

\bibitem{Vuillot-arxiv19}
\BIBentryALTinterwordspacing
C.~Vuillot and N.~P. Breuckmann, ``{Quantum Pin Codes},'' \emph{arXiv preprint
  arXiv:1906.11394}, 2019. [Online]. Available:
  \url{http://arxiv.org/abs/1906.11394}
\BIBentrySTDinterwordspacing

\bibitem{Rengaswamy-isit18}
\BIBentryALTinterwordspacing
N.~Rengaswamy, R.~Calderbank, S.~Kadhe, and H.~D. Pfister, ``Synthesis of
  logical {C}lifford operators via symplectic geometry,'' in \emph{Proc.\ IEEE
  Int.\ Symp.\ Inform.\ Theory}.\hskip 1em plus 0.5em minus 0.4em\relax IEEE,
  2018, pp. 791--795. [Online]. Available:
  \url{http://arxiv.org/abs/1803.06987}
\BIBentrySTDinterwordspacing

\bibitem{Boykin-arxiv99}
P.~O. Boykin, T.~Mor, M.~Pulver, V.~Roychowdhury, and F.~Vatan, ``{On Universal
  and Fault-Tolerant Quantum Computing},'' \emph{arXiv preprint
  arXiv:quant-ph/9906054}, 1999, [Online]. Available:
  http://arxiv.org/abs/quant-ph/9906054.

\bibitem{Gottesman-nature99}
\BIBentryALTinterwordspacing
D.~Gottesman and I.~L. Chuang, ``{Demonstrating the viability of universal
  quantum computation using teleportation and single-qubit operations},''
  \emph{Nature}, vol. 402, no. 6760, pp. 390--393, 1999. [Online]. Available:
  \url{http://www.nature.com/articles/46503}
\BIBentrySTDinterwordspacing

\bibitem{Cui-physreva17}
\BIBentryALTinterwordspacing
S.~X. Cui, D.~Gottesman, and A.~Krishna, ``{Diagonal gates in the Clifford
  hierarchy},'' \emph{Phys. Rev. A}, vol.~95, no.~1, p. 012329, 2017, [Online].
  Available: http://arxiv.org/abs/1608.06596. [Online]. Available:
  \url{https://journals.aps.org/pra/pdf/10.1103/PhysRevA.95.012329}
\BIBentrySTDinterwordspacing

\bibitem{Rengaswamy-pra19}
\BIBentryALTinterwordspacing
N.~Rengaswamy, R.~Calderbank, and H.~D. Pfister, ``Unifying the {C}lifford
  hierarchy via symmetric matrices over rings,'' \emph{Phys. Rev. A}, vol. 100,
  no.~2, p. 022304, 2019. [Online]. Available:
  \url{http://arxiv.org/abs/1902.04022}
\BIBentrySTDinterwordspacing

\bibitem{Wilde-physreva09}
\BIBentryALTinterwordspacing
M.~M. Wilde, ``{Logical operators of quantum codes},'' \emph{Phys. Rev. A},
  vol.~79, no.~6, p. 062322, 2009. [Online]. Available:
  \url{https://arxiv.org/abs/0903.5256}
\BIBentrySTDinterwordspacing

\bibitem{Rengaswamy-arxiv19b}
\BIBentryALTinterwordspacing
N.~Rengaswamy, R.~Calderbank, S.~Kadhe, and H.~D. Pfister, ``{Logical Clifford
  Synthesis for Stabilizer Codes},'' \emph{arXiv preprint arXiv:1907.00310},
  2019. [Online]. Available: \url{http://arxiv.org/abs/1907.00310}
\BIBentrySTDinterwordspacing

\bibitem{Randriambololona-aagct15}
H.~Randriambololona, ``On products and powers of linear codes under
  componentwise multiplication,'' \emph{Algorithmic arithmetic, geometry, and
  coding theory}, vol. 637, pp. 3--78, 2015.

\bibitem{Campbell-blog16}
\BIBentryALTinterwordspacing
E.~T. Campbell, ``The smallest interesting colour code,'' 2016, blog post.
  [Online]. Available:
  \url{https://earltcampbell.com/2016/09/26/the-smallest-interesting-colour-code/}
\BIBentrySTDinterwordspacing

\bibitem{Bardet-isit16}
\BIBentryALTinterwordspacing
M.~Bardet, V.~Dragoi, A.~Otmani, and J.-P. Tillich, ``Algebraic properties of
  polar codes from a new polynomial formalism,'' in \emph{Proc.\ IEEE Int.\
  Symp.\ Inform.\ Theory}.\hskip 1em plus 0.5em minus 0.4em\relax IEEE, 2016,
  pp. 230--234. [Online]. Available: \url{http://arxiv.org/abs/1601.06215}
\BIBentrySTDinterwordspacing

\bibitem{Krishna-arxiv18}
\BIBentryALTinterwordspacing
A.~Krishna and J.-P. Tillich, ``Magic state distillation with punctured polar
  codes,'' \emph{arXiv preprint arXiv:1811.03112}, 2018. [Online]. Available:
  \url{http://arxiv.org/abs/1811.03112}
\BIBentrySTDinterwordspacing

\bibitem{Ax-ajm64}
\BIBentryALTinterwordspacing
J.~Ax, ``{Zeroes of Polynomials Over Finite Fields},'' \emph{Am. J. Math.},
  vol.~86, no.~2, p. 255, Apr 1964. [Online]. Available:
  \url{https://www.jstor.org/stable/2373163?origin=crossref}
\BIBentrySTDinterwordspacing

\end{thebibliography}
\end{document}